\documentclass{ws-ijmpa}

\begin{document}

\markboth{T. R. P. Caram\^es, E. R. Bezerra de Mello \& M. E. X. Guimar\~aes}
{Gravitational Field of a Global Monopole in a Modified Gravity}

%
\catchline{}{}{}{}{}
%

\title{GRAVITATIONAL FIELD OF A GLOBAL MONOPOLE IN A MODIFIED GRAVITY}

\author{T. R. P. CARAM\^ES}

\address{Instituto de F\'isica, Universidade Federal Fluminense, Av. Gal. Milton Tavares de Souza, s/nº - Campus da Praia Vermelha - Niter\'oi - Rio de Janeiro/ CEP 24210-346, Brazil\\
carames@if.uff.br}

\author{E. R. BEZERRA DE MELLO}

\address{Departamento de F\'isica, Universidade Federal da Para\'iba, Cidade Universit\'aria\\
Jo\~ao Pessoa, Para\'iba/ CEP 58051-900, Brazil\\
emello@fisica.ufpb.br}

\author{M. E. X. GUIMAR\~AES}

\address{Instituto de F\'isica, Universidade Federal Fluminense, Av. Gal. Milton Tavares de Souza, s/nº - Campus da Praia Vermelha - Niter\'oi - Rio de Janeiro/ CEP 24210-346, Brazil\\
emilia@if.uff.br}

\maketitle

\begin{history}
\received{Day Month Year}
\revised{Day Month Year}
\end{history}

\begin{abstract}
In this paper we analyze the gravitational field of a global monopole in the context of $f(R)$ gravity. More precisely, we show that the field equations obtained are expressed in terms of $F(R)=\frac{df(R)}{dR}$. Since we are dealing with a spherically symmetric system, we assume that $F(R)$ is a function of the radial coordinate only. Moreover, adopting the weak field approximation, we can provide all components of the metric tensor. A comparison with the corresponding results obtained in General Relativity and in the Brans-Dicke theory is also made.

\end{abstract}

\ccode{PACS numbers: $04.50.+h$, $04.20.-q$, $14.80.Hv$}

\section{Introduction}

It is well known that several types of topological defects may have been created
by the vacuum phase transitions in the early universe.\cite{kibble,vilenkin}
These include domain walls, cosmic strings and monopoles. Global monopoles are
heavy objects formed in the phase transition of a system composed by a self-coupling
triplet of scalar fields $\phi^{a}$ ($a=1,2,3$) whose original global $O(3)$ symmetry
is spontaneously broken to $U(1)$. The scalar matter field plays the role of an order
 parameter which outside the monopole's core acquires a non-vanishing value.
 The spacetime associated with global monopoles is characterized by a non-trivial
 topology observed as a deficit solid angle. The properties of the gravitational
 field produced by this exotic object have been investigated by M. Barriola and
  A. Vilenkin in the context of the general relativity.\cite{barriola} Later,
  the gravitational field of a global monopoles were analyzed by A. Barros and
  C. Romero in Ref.~\refcite{romero} in the Brans-Dicke theory of gravity in the
  context of weak field approximation.

On the other hand, the $f(R)$ theories of gravity have been recently suggested
as a possible alternative to explain the late time cosmic speed-up experienced
by our universe.\cite{Carrol,Fay} Such theories avoid the Ostrogradski's instability
that can otherwise prove to be problematic for general higher derivatives theories.\cite{Ostr,Woo}

In this work we analyse the gravitational field of the global monopole in the $f(R)$
theories of gravity scenario, handling the field equations at the weak field regime.
In order to simplify the form of the fields equations, we shall express them in terms
of the function $F(R)=\frac{df(R)}{dR}$, in a similar procedure as adopted in
Refs.~\refcite{Mut} and ~\refcite{carames}. Since we are dealing with a spherically
symmetric static system, we shall express $F(R)$ as a function of the radial coordinates,
$r$, by defining $F(R(r))={\cal F}(r)$.

\section{The Global Monopole Model in General Relativity}
\label{sec2}

The simplest model which gives rise to a global monopole is described by the
Lagrangian density below:\cite{barriola}
\begin{equation}
\label{lagrangian}
{\cal L}=\frac{1}{2}\partial_{\mu}\phi^a \partial^{\mu}\phi^a-
\frac{1}{4} \lambda (\phi^{a}\phi^{a}-\eta^2)^2\ .
\end{equation}
Coupling this matter field with the Einstein equations, spherically symmetric
solutions for the fields equations can be obtained by adopting for the line element
and the matter fields the {\it Ans\"atze} below:
\begin{eqnarray}
\label{sph}
ds^2 =B(r)dt^2-A(r)dr^2-r^2(d\theta^2 + \sin^2\theta d\varphi^2)
\end{eqnarray}
and
\begin{equation}
\label{campo}
\phi^{a}=\eta h(r) \frac{x^{a}}{r}
\end{equation}
with $x^ax^a=r^2$.

For points far from the monopole's core, the solutions for the functions $B(r)$
and $A(r)$ are given by
\begin{eqnarray}
B = A^{-1}\approx 1 - 8\pi G\eta^2 - 2GM/r \ ,
\end{eqnarray}
being the mass parameter $M\approx M_{core}$. What concerns the matter field,
we have $h(r)\approx 1$. Moreover, in Ref.~\refcite{barriola} the authors have shown
that for points outside the monopole's core, the energy-momentum tensor associated
with a static global monopole can be approximated as
\begin{equation}
\label{emt}
T_\nu^\mu\approx{\rm diag}\left(\frac{\eta^2}{r^2},\frac{\eta^2}{r^2},0,0\right)\ .
\end{equation}

\section{The $f(R)$ Gravity in the Metric Formalism}

The action associated with the modified theories of gravity coupled with matter fields is given by:
\begin{equation}
\label{Act}
S=\frac1{2\kappa}\int d^{4}x\sqrt{-g}f(R)+{\cal S}_m \ ,
\end{equation}
where $f(R)$ is an analytical function of the the Ricci scalar, $R$,
$\kappa=8\pi G$ and ${\cal S}_m$ corresponds to the action associated with the matter fields.
By using the metric formalism, the field equations become:
\begin{equation}
\label{G}
G_{\mu\nu}\equiv R_{\mu\nu}-\frac{1}{2}Rg_{\mu\nu}=T^{c}_{\mu\nu}+\kappa \tilde{T}^{m}_{\mu\nu} \ ,
\end{equation}
in which $T^{c}_{\mu\nu}$ is the geometric energy-momentum tensor, namely
\begin{eqnarray}
\label{3}
T^{c}_{\mu\nu}=\frac{1}{F(R)}\left\{\frac{1}{2}g_{\mu\nu}\left(f(R)-F(R)R\right)+
\nabla^{\alpha} \nabla^{\beta} F(R)\left(g_{\alpha\mu}g_{\beta\nu}-
g_{\mu\nu}g_{\alpha\beta}\right)\right\}
\end{eqnarray}
with $F(R)\equiv\frac{df(R)}{dR}$.

The standard minimally coupled energy-momentum tensor, ${T}^{m}_{\mu\nu}$, derived from
the matter action, is related to $\tilde{T}^m_{\mu\nu}$ by
\begin{equation}
\label{T}
\tilde{T}^{m}_{\mu\nu}=T^m_{\mu\nu}/F(R) \ .
\end{equation}
Thus the field equations can be written as
\begin{eqnarray}
\label{FE}
F(R)R_{\mu\nu}-\frac{1}{2}f(R)g_{\mu\nu}-\nabla_{\mu}\nabla_{\nu}F(R)+g_{\mu\nu}\Box F(R)=
\kappa T^{m}_{\mu\nu}\ .
\end{eqnarray}
Taking the trace of the above equation we get
\begin{equation}
\label{6}
F(R)R-2f(R)+3\Box{F(R)}=\kappa T^{m} \ ,
\end{equation}
which expresses a further scalar degree of freedom that arises in the modified theory.
Through this equation it is possible to express $f(R)$ in terms of its derivatives and
the trace of the matter energy-momentum tensor as follows:
\begin{eqnarray}
\label{fr}
f(R)=\frac{1}{2}\left(F(R)R+3\Box{F(R)}-\kappa T^{m}\right) \ .
\end{eqnarray}
Substituting the above expression into (\ref{FE}) we obtain
\begin{align}
F(R)R_{\mu\nu}-\nabla_{\mu}\nabla_{\nu}F(R)-\kappa \tilde{T}^{m}_{\mu\nu}=
\frac{g_{\mu\nu}}{4}\left[F(R)R-\Box F(R)-\kappa T^{m}\right] \ .
\end{align}
From this expression we can see that the combination below
\begin{align}
\label{A0}
    C_\mu=\frac{F(R)R_{\mu\mu}-\nabla_{\mu}\nabla_{\mu}F(R)-\kappa T^{m}_{\mu\mu}}{g_{\mu\mu}} \ ,
\end{align}
with fixed indices, is independent of the corresponding index. So, the following relation
\begin{equation}
\label{Ceq}
C_{\mu}-C_{\nu}=0\ ,
\end{equation}
holds for all $\mu$ and $\nu$.

\section{The Global Monopole in $f(R)$ Theories}
\label{sec3}
In this section we shall derive the field equations associated with a global monopole
system in modified theories of gravity.

By adopting a spherical symmetry {\it Anstaz} for the metric tensor given in (\ref{sph}),
the non-vanishing components of the Ricci tensor read:
\begin{equation}
\label{00}
R_0^0=-\frac{1}{2}\left[\frac{(B')^{2}}{2B^2A}+\frac{B'A'}{2BA^{2}}-\frac{B''}{BA}-
\frac{2B'}{BAr}\right],
\end{equation}
\begin{equation}
\label{11}
R_1^1=-\frac{1}{2}\left[\frac{B'A'}{2BA^2}+\frac{2A'}{rA^2}-\frac{B''}{BA}+
\frac{(B')^{2}}{2B^{2}A}\right],
\end{equation}
\begin{equation}
\label{22}
R_2^2=-\frac{A'}{2rA^{2}}+\frac{B'}{2rBA}+\frac{1}{r^2A}-\frac{1}{r^2}\ .
\end{equation}
The spherical symmetry requires that
\begin{equation}
R^{2}_{2}=R^{3}_{3},
\end{equation}
which implies that the scalar curvature will be given by
\begin{eqnarray}
\label{Ricci}
R=\frac{2}{r^{2}A}\left\{1-A+\left(\frac{B'}{B}-\frac{A'}{A}\right)r-
\frac{1}{4}\frac{B'}{B}r^{2}\left(\frac{B'}{B}+\frac{A'}{A}\right)+
\frac{1}{2}r^{2}\frac{B''}{B}\right\} \ .
\end{eqnarray}
In all the above equations the primes corresponds derivative of the function with
respect to the radial coordinate. By using the identity (\ref{Ceq}) along with the
energy-momentum tensor given by (\ref{emt}), we can construct two linearly independent
differential equations:
\begin{equation}
\label{Y1}
    2r{\cal F}^{''}-r{\cal F}'\left(\frac{B'}{B}+\frac{A'}{A}\right)-
    2{\cal F}\left(\frac{B'}{B}+\frac{A'}{A}\right)=0 \ ,
\end{equation}
and
\begin{eqnarray}
\label{Y2}
&-&4B+4AB-4rB\frac{{\cal F}'}{{\cal F}}+2r^{2}B'\frac{{\cal F}'}{{\cal F}}-
r^{2}B'\left(\frac{B'}{B} +\frac{A'}{A}\right)\nonumber\\
&+&2r^{2}B^{''}+2Br\left(\frac{B'}{B}+\frac{A'}{A}\right)-\frac{4AB \kappa \eta^2}{\cal F}=0\ ,
\end{eqnarray}
where we have expressed $F(R)$ as $F(R(r))={\cal F}(r)$.

Defining
\begin{equation}
\label{beta2}
\beta \equiv \frac{B'}{B}+\frac{A'}{A}\ ,
\end{equation}
we can write the field equations as follows
\begin{equation}
\label{beta}
\frac{\beta}{r}=\frac{{\cal F}''}{{\cal F}}-\frac{1}{2}\frac{{\cal F}'}{{\cal F}}\beta\ ,
\end{equation}
and
\begin{eqnarray}
&-&4B+4AB-4rB\frac{{\cal F}'}{{\cal F}}+2r^{2}B'\frac{{\cal F}'}{{\cal F}} \nonumber\\
&+&2r^{2}B^{''}-r^2B'\beta+2Br\beta-\frac{4AB\kappa \eta^2}{{\cal F}}=0\ .
\end{eqnarray}
We can verify that if we plug $\omega=0$ into the equation $(9)$ of Ref.~\refcite{romero}
and  replace $\phi(r)$ by ${\cal F}(r)$, the equation (\ref{beta}) is recovered.
This emphasizes the equivalence of the metric $f(R)$ gravity with a Brans-Dicke gravity
possessing a parameter $\omega=0$, as pointed out in Ref.~\refcite{faraoni}. Furthermore
this fact shows the relevance of working with the fields equations in terms of $F(R)$ when
one wants to compare with the results obtained via Brans-Dicke gravity, since the role of
the scalar degree of freedom in $f(R)$ gravity is played by the function $F(R)$.

\section{Solutions in the Weak Field Regime}

Now let us consider the weak field approximation in the field equations by assuming
that $B(r)=1+b(r)$ and $A(r)=1+a(r)$ with $\left|b(r)\right|$ and $\left|a(r)\right|$
smaller than unity. Furthermore, let us consider that the modification theory of gravity
corresponds to a small correction on General Relativity (GR), so that
${\cal F}(r)=1+\psi(r)$, with $\left|\psi(r) \right|<<1$

Adopting these approximations it is possible to verify that up to the first order we may write,
\begin{eqnarray}
\label{app}
&&\frac{{\cal F}'}{{\cal F}}=\frac{\psi'}{1+\psi}\approx\psi'\;\; ,
\frac{{\cal F}''}{{\cal F}}=\frac{\psi''}{1+\psi}\approx\psi''\ , \nonumber \\
&&\frac{B'}{B}=\frac{b'}{1+b}\approx b'\;\; ,\frac{A'}{A}=\frac{a'}{1+a}\approx a'\ .
\end{eqnarray}

Thus the approximated equations for ($\ref{Y1}$) and ($\ref{Y2}$) read
\begin{equation}
\label{ap1}
\frac{\beta}{r}=\psi''
\end{equation}
and
\begin{eqnarray}
\label{ap2}
&&4a-4r\psi'+2r(a'+b')-2r^2b''-4(1+a+b-\psi)\kappa \eta^2=0\ .
\end{eqnarray}
Specific results for the fields equations can be obtained by adopting for $\psi(r)$ a
specific {\it Ansatz}. Considering the simplest analytical function of the radial
coordinate, namely $\psi(r)=\psi_0 r$, we shall have the following result for Eq. ($\ref{ap1}$):
\begin{eqnarray}
\label{b1}
\frac{\beta}{r}=0\ ,
\end{eqnarray}
which means
\begin{eqnarray}
a'+b'=0 \\
a+b=c_0 \ ,
\end{eqnarray}
where $c_0$ is an integration constant that we choose as $c_0=0$. Therefore
\begin{equation}
a(r)=-b(r)\ .
\end{equation}
So (\ref{ap2}) will be written as
\begin{equation}
\frac{1}{2}r^2b''-b-r\psi_0-(1-\psi_0 r)\kappa \eta^2=0\ ,
\end{equation}
whose solution is
\begin{equation}
b(r)=\frac{c_1}{r}+c_2r^2-\kappa \eta^2-\psi_0 r(1-\kappa \eta^2)\ ,
\end{equation}
where $c_1$ and $c_2$ are integration constants. Since we are not dealing with a theory
with a cosmological constant we may set $c_2=0$. Furthermore, we must choose $c_1=-2GM$
in order to recover the Newtonian potential present in the Barriola and Vilenkin solution.
Therefore,
\begin{equation}
B(r)=1-\frac{2GM}{r}-8\pi G\eta^2-\psi_0 r(1-8\pi G \eta^2)\ ,
\end{equation}
since $\kappa=8 \pi G$. For a typical Grand Unified Theory the parameter $\eta$ is of
the order $10^{16}$ Gev. So, $8\pi G \eta^2\approx 10^{-5}$.\cite{vilenkin} This allows us
to neglect the term $\psi_0 r\times 8\pi G \eta^2$ in the above result since
$\left|\psi_0 r\right|<<1 $. Then we may write
\begin{equation}
\label{sol1}
B(r)=1-\frac{2GM}{r}-8\pi G\eta^2-\psi_0 r \ .
\end{equation}
From equation ($\ref{b1}$) we have
\begin{equation}
A(r)=\frac{a_0}{B(r)}\ ,
\end{equation}
where $a_0$ is an integration constant. Rescaling the time coordinate we can set $a_0=1$. Then,
\begin{equation}
\label{sol2}
A(r)=\left[1-\frac{2GM}{r}-8\pi G\eta^2-\psi_0 r\right]^{-1}\
\end{equation}
As previously assumed by Barriola and Vilenkin, here we also drop out the mass term
in (\ref{sol1}) and (\ref{sol2}) as it is negligibly small at astrophysical scale.
Thus, we have
\begin{equation}
\label{sola}
B(r)=1-8\pi G\eta^2-\psi_0 r\
\end{equation}
and
\begin{equation}
\label{solb}
A(r)=\left(1-8\pi G\eta^2-\psi_0 r\right)^{-1}\ .
\end{equation}
Using the weak field approximation in (\ref{Ricci}) we can obtain the Ricci scalar
curvature associated with the solution above:
\begin{equation}
\label{RicciM}
R=-\frac{16\pi G \eta^2}{r^2}-\frac{6\psi_0}{r}\ .
\end{equation}
From ${\cal F}(r)$ and $R(r)$ one can determine, in principle, $F(R)$ and finally $f(R)$.
The explicit form obtained for $f(R)$ is given by
\begin{eqnarray}
\label{fR}
&&f(R)=R-3\psi_0^2 \ln \left(\frac{R}{R_0}\right)-
2 \psi_0 \sqrt{9 \psi_0^2-16 \pi G \eta^2 R}\nonumber\\
&-&3 \psi_0^2 \ln \left[ \frac{\sqrt{9 \psi_0^2-16 \pi G \eta^2 R}
-3 \psi_0}{\sqrt{9 \psi_0^2-16 \pi G \eta^2 R_0}-3 \psi_0}\right]+
3 \psi_0^2 \ln \left[\frac{\sqrt{9 \psi_0^2-16 \pi G \eta^2 R}+
3 \psi_0}{\sqrt{9 \psi_0^2-16 \pi G \eta^2 R_0}+3 \psi_0}\right]\ ,\nonumber\\
\end{eqnarray}
where $R_0$ is a constant. It is straightforward to verify that if $\psi_0$ is set to
be positive, the function above satisfies the following stability conditions required for
any physically relevant $f(R)$ theory:\cite{pogosian}\cdash\cite{faraoni}
\begin{itemize}
\item $\frac{d^2f(R)}{dR^2}>0$ (no tachyons);
\item $\frac{df(R)}{dR}>0$ (no ghosts);
\item $\lim_{R \rightarrow \infty} \frac{\Delta}{R}=0$ and $\lim_{R \rightarrow \infty}
\frac{d \Delta}{dR}=0$ (GR is recovered at early times),
\end{itemize}
where $\Delta=\Delta(R)$ is defined as $\Delta=f(R)-R$.

Now we shall show that the line element described by the obtained functions $A(r)$
and $B(r)$ are conformally related to the ordinary global monopole
solutions.\footnote{An analogue procedure has been adopted in Ref.~\refcite{romero}.}
So let us consider the following coordinate transformation:
\begin{eqnarray}
B(r)&=&p(r^{*})\left(1-8\pi G \eta^2\right) \ , \\
\label{transfA}
A(r)dr^2&=&p(r^{*})\left(1+8 \pi G \eta^2\right)(dr^{*})^2 \ , \\
r&=&p^{1/2}(r^*)r^* \ ,
\label{trans}
\end{eqnarray}
where $h(r^*)$ is an arbitrary function of $r^*$ to be determined, and $h(r^*)=
1+q(r^*)$ with $\left|q(r^*)\right|<<1$.
Differentiating equation (\ref{trans}) we have
\begin{equation}
\label{eqpq}
dr^2=\left(1+r^*\frac{dq}{dr^*}+q\right)dr^{*2}\ .
\end{equation}

Substituting the above equation into (\ref{transfA}) and keeping only  linear terms
in $q(r^*)$, $\psi_0r$ and $G\eta^2$, we obtain the following result for $q(r^{*})$:
\begin{equation}
q(r^*)=-\psi_0r^*\ ,
\end{equation}
then
\begin{equation}
p(r^*)=1-\psi_0r^*\ .
\end{equation}

Thus we can write the line element (\ref{sph}) in the coordinate $r^*$ as follows:
\begin{eqnarray}
\label{line1}
ds^2&=&\left(1-\psi_0r^*\right)\left[\left(1-8\pi G \eta^2\right)dt^2-
\left(1+8\pi G \eta^2\right)dr^{*2}\right.\nonumber\\
&-&\left.r^{*2}\left(d\theta^2+\sin^2\theta d\phi^2 \right)\right]\ .
\end{eqnarray}
Rescaling the time coordinate and redefining the radial coordinate as
$r=\left(1+4 \pi G \eta^2 \right)r^*$, we arrive at the line element below:
\begin{eqnarray}
\label{mgm}
ds^2&=&\left(1-\psi_0r\right)\left[dt^2-dr^2-
\left(1-8 \pi G \eta^2 \right)r^2\left(d\theta^2+\sin^2\theta d\phi^2 \right)\right]\ .
\end{eqnarray}

An important feature arises if we are interested in analyzing the deflection of light in
this metric. As it is well known, the deflection angles are always preserved for two metrics
related by a conformal transformation. Therefore, the deflection of a light ray by the
monopole in the present modified gravity will be the same of that one previously obtained
in Refs.~\refcite{barriola}:
\begin{equation}
\delta \phi = 8 \pi G \eta^{2} l(d+l)^{-1}\ ,
\end{equation}
where $d$ and $l$  are the distances from the monopole to the observer  and to the source,
respectively. In this analysis it was assumed that the light ray trajectory lies at the
equatorial plane $\theta= \frac{\pi}{2}$.
\\
\section{Conclusions}

In this work we have analyzed the gravitational field of a global monopole in the
$f(R)$ gravity scenario in the metric formalism. In this formalism, this modified gravity contains a massive scalar
degree of freedom in addition to the familiar massless graviton and it turns
out to be equivalent to a Brans-Dicke theory. In order to simplify our analysis
we have considered solutions in the weak field approximation which implied that we are considering a theory which is 
as a small correction on GR. The latter condition was explicitly considered by assuming
${\cal{F}}(r)=1+\psi_0r$, for $|\psi_0r|<<1$.

Following the above mentioned approach, the solutions found by us correspond to small
corrections on $g_{00}$ and $g_{11}$ components of the metric tensor, only. Being these
new components given in (\ref{sol1}) and (\ref{sol2}), respectively. From the results obtained,
we can observe that also, a small correction on the Ricci scalar, given now by (\ref{RicciM}),
takes place. Moreover, we have also verified that these solutions are conformally
related to that one previously obtained by M. Barriola and A. Vilenkin in
Ref.~\refcite{barriola} as shown in (\ref{mgm}), what ensures us that the deflection
of light in these two spacetimes will be the same.

Before finishing this section, we would like to point out that by using the weak field approximation,
we were able to provide an explicit expression for the function $f(R)$, given in
(\ref{fR}). The non-linear function $f(R)$ is not a problem if we consider low curvatures (i.e., late time in the matter dominated era). Also, by assuming that $\psi_0$ is a positive parameter, we can affirm that $f(R)$ fulfills two important stability conditions, namely, non-tachyons and non-ghosts requirements. 

\section{Acknowledgments}

TRPC thanks CAPES for financial support and Prof. S. Jor\'as for fruitful discussions
and useful suggestions. ERBM and MEXG thank the Conselho Nacional de Desenvolvimento
 Cient\'\i fico e Tecnol\'ogico (CNPq) for partial financial support.

\end{document}